\documentclass{pasj01}
\usepackage{subfigure}

\Received{$\langle$reception date$\rangle$}
\Accepted{$\langle$acception date$\rangle$}
\Published{$\langle$publication date$\rangle$}

\begin{document}

\title{MAXI upper limits of the electromagnetic counterpart of GW170817}

\author{Satoshi~\textsc{Sugita}\altaffilmark{1,2}}%
\email{sugita@phys.aoyama.ac.jp}
\author{Nobuyuki~\textsc{Kawai}\altaffilmark{1}}%
\author{Satoshi~\textsc{Nakahira}\altaffilmark{3}}%
\author{Hitoshi~\textsc{Negoro}\altaffilmark{4}}%
\author{Motoko~\textsc{Serino}\altaffilmark{2}}%
\author{Tatehiro~\textsc{Mihara}\altaffilmark{3}}%
\author{Kazutaka~\textsc{Yamaoka}\altaffilmark{5}}%
\author{Motoki~\textsc{Nakajima}\altaffilmark{6}}%

\altaffiltext{1}{Department of Physics, Tokyo Institute of Technology, 
2-12-1 Ookayama, Meguro-ku, Tokyo 152-8551, Japan}
\altaffiltext{2}{College of Science and Engineering, Department of Physics and Mathematics, Aoyama Gakuin University, 5-10-1 Fuchinobe, Chuo-ku, Sagamihara, Kanagawa 252-5258, Japan}
\altaffiltext{3}{MAXI team, RIKEN, 2-1 Hirosawa, Wako, Saitama 351-0198, Japan}
\altaffiltext{4}{Department of Physics, Nihon University, 
1-8-14 Kanda-Surugadai, Chiyoda-ku, Tokyo 101-8308, Japan}
\altaffiltext{5}{Institute for Space-Earth Environmental Research (ISEE), Nagoya University, Furo-cho, Chikusa-ku, Nagoya, Aichi 464-8601, Japan}
\altaffiltext{6}{School of Dentistry at Matsudo, Nihon University, 2-870-1, Sakaecho-nishi, Matsudo, Chiba 271-8587, Japan}


\maketitle

\begin{abstract}

We report the MAXI observation of the gravitational-wave (GW) event GW170817 
and the electromagnetic counterpart of GW170817.
GW170817 is a binary neutron star coalescence candidate 
detected by the Advanced LIGO and Advanced Virgo detectors,  
and it is the first event for which the optical counterpart has been discovered.
In the MAXI observation, 
the Gas Slit Camera (GSC) covered approximately 62\% of the sky region of the GW event within 90\% probability during the first 92 min of orbit after the trigger.
No significant X-ray transient was detected in the error region, and 
the upper limit of the average flux with a significance of 3 $\sigma$ in the 2--10 keV band was 53/26 mCrab (one-orbit observation/one-day observation).
In the optical counterpart of GW170817, 
the observational window of GSC at the position started at 20 s after the GW trigger, 
but the high voltage of GSC was unfortunately off at the time because the ISS was entering a high-particle-background region.
The first observation of the position by GSC was eventually performed at 16797 sec (4.6 hours) since the GW trigger, 
yielding the 3 $\sigma$ upper limit of 8.60$\times$10$^{-9}$ erg cm$^{-2}$ s$^{-1}$ in the 2--10 keV band, 
though it was the earliest X-ray observation of the counterpart.

\end{abstract}

\section{Introduction}

The Advanced Laser Interferometer Gravitational-wave Observatory (LIGO) and the Advanced Virgo detectors
observed a gravitational-wave signal named GW170817.
The signal was consistent with a binary neutron star (BNS) coalescence with a merger time of August 17, 2017 12:41:04 UTC \citep{2017GCN.21509}. 
The sky position of GW170817 was localized to an area of 28 deg$^{2}$ 
with a centroid of (RA, Dec) = (\timeform{197D.25}, \timeform{-25D.62})  \citep{2017GCN.21513,2017GCN.21527}.
The luminosity distance to the source was 40$^{+8}_{-14}$ Mpc,
making GW170817  the closest GW event ever observed.
From the waveform analysis, 
the masses of the BNS were estimated to be in the range of $m_{1} = (1.36\mathchar`-2.26) M_{\odot}$ 
and $m_{2} = (0.86\mathchar`-1.36)M_{\odot}$ \citep{PhysRevLett.119.161101}. 
\\
On August 17 2017 12:41:06 UTC, 
the Fermi Gamma-ray Burst Monitor (GBM) and the Anti-Coincidence Shield (ACS) of the International Gamma-Ray Astrophysics Laboratory (INTEGRAL) detected a short gamma-ray burst (GRB) named GRB170817A \citep{2017GCN.21520, 2017ApJ...848L..15S}, which was weak and had a duration of  $2$ s.
The trigger time of GRB170817A was 1.7 s after the trigger time of GW170817,
and its position was localized to (RA, Dec) = (\timeform{176D.8}, \timeform{-39D.8}) with a 90\% probability region of approximately 1800 deg$^{2}$, 
which falls within the error region of GW170817. 
\\
The above results imply that GRB170817A was associated with the BNS merger candidate GW170817.
The three-dimensional localization of GW170817 and the detection of GRB170817A were reported by the Gamma-ray Coordinates Network (GCN) to electromagnetic observation teams around the world.
A large number of the teams performed observations to find the electromagnetic counterpart of the BNS merger.
The first report to GCN on the discovery was the observation by the Swope telescope at Las Campanas Observatory in Chile \citep{2017GCN.21529}.
The observation was performed on August 17 23:33 UTC, 10.87 h from the GW trigger.
The transient, named Swope Supernova Survey 2017a (SSS17a), was i = 17.476 $\pm$ 0.018 mag, 
and localized at (RA,Dec) = (\timeform{197D.45},\timeform{-23D.38}).
SSS17a was located in NGC 4993 with an offset of 10.6 arc sec (corresponding to 2.0 kpc at 40 Mpc) from the center of NGC 4993 \citep{Coultereaap9811}.
NGC 4993 was firmly located inside the 3D skymap of GW170817; therefore, it is most likely the host galaxy of GW170817.
A large number of multi-wavelength follow-up observations were performed for the position of SSS17a \citep{2017ApJ...848L..12A}.
In X-ray observation, Monitor of All-sky X-ray Image (MAXI) and the Super--AGILE onboard AGILE observed the position of the optical counterpart and reported the upper limit of X-ray flux \citep{2017GCN.21555}.
After the confirmation of the position of the counterpart, 
pointing X-ray observations were performed to searching for X-ray afterglow from GRB170817A/GW170817
by using Swift-XRT \citep{2017GCN.21550,2017GCN.21612}, $NuSTAR$ \citep{2017GCN.21626}, INTEGRAL JEM-X \citep{2017GCN.21672}, and Chandra \citep{2017GCN.21765}.
In Chandra, the first follow-up observation conducted 2.3 days after the GW trigger did not detect any X-ray source, 
but the second observation starting 9 days after the trigger
detected an X-ray source, the position of which was consistent with SSS17a \citep{2017Natur.551...71T, 2017ApJ...848L..20M}.
\\
MAXI \citep{2009PASJ...61..999M} is a mission onboard the Japanese Experimental Module-Exposed Facility (JEM-EF) on the International Space Station (ISS).
MAXI scans $\sim$85\% of the whole sky in one orbit (92 min) by sweeping with a slit-shaped field of view (FOV).
It can cover a large localization area of a GW event detected by GW detectors 
and search for an emission from the area before the time of GW trigger.
From the operation start of LIGO, MAXI had searched for X-ray counterparts of GW events 
and reported upper limits of X-ray flux in the localization areas of GW150914 \citep{2017PASJ...69...84K}, GW151226 \citep{2017PASJ...69...85S}, 
GW170104 \citep{2017GCN.20507}, and GW170814 \citep{2017GCN.21494}.
In this paper, we present detailed results of the MAXI observation of GW170817 and its electromagnetic counterpart, 
following a quick report of X-ray upper limits from MAXI observation.

\section{Observation}

\subsection{Instrumentation and Operation}

MAXI on ISS has two instruments: the Gas Slit Camera (GSC) \citep{2011PASJ...63S.623M}
and the Solid-state Slit Camera (SSC) \citep{2011PASJ...63..397T}.
The GSC does not operate in regions with a high particle background, 
including the South Atlantic Anomaly; regions with a 
latitude higher than $\sim$40 deg; 
and in an FOV around the sun ($\sim5$ deg).
Although its operating duty ratio is approximately 40\%,
the GSC covers approximately 85 \% of the whole sky in a scan
\citep{2011PASJ...63S.635S}.
Because the SSC is operated in the night time to avoid sunlight,
the SSC duty ratio and sky coverage are approximately 25-30\% and 30\%, respectively. 
In the observation of GW170817, the 90\% probability region observed by LIGO/Virgo was out of the FOV of SSC.
\\
The GSC system is composed of 12 cameras.
Each camera consists of a slit-slat collimator and proportional counter with one-dimensional position sensitivity, covering an energy range of 2-30 keV. 
Six of the twelve cameras are assembled into two modules, which cover a wide rectangular FOV of 1.5 (FWHM)$\times$160 deg (orbital$\times$orthogonal direction).
The FOV of one module points toward the Earth horizon 
and that of the other module points toward the Earth zenith, 
covering approximately 2\% of the whole sky.
A scanning image of an object is obtained with the triangular response of the slat collimator according to the ISS orbital motion. 
Each photon direction typically has errors of 1.5 deg in the scan direction and $\sim$2 deg in the orthogonal direction at FWHM, 
corresponding to the point spread function (PSF) of the camera. 
A nominal GSC camera can detect transient events with 
a 2--20 keV flux greater than 2 $\times 10^{-9}$ erg cm$^{-2}$ s$^{-1}$ 
(e.g., \cite{2014PASJ...66...87S,2016PASJ...68S...1N}) in a scan transit.
At the time of the GW170817 observation, seven cameras of ID 0, 2, 3, 4, 5, 6, and 7 were functioning \citep{2014SPIE.9144E..1OM},  
and the GSC observed the position using camera ID 2 on the zenith module and  using camera ID 4 and 5 on the horizontal module for 10 days after the GW trigger.

\subsection{Observation of GW170817}

\begin{figure}[htp]
 \begin{center}
 \subfigure[one-orbit observation (from T0+0 to T0+5520 sec)]{
 \includegraphics[width=6.5cm] {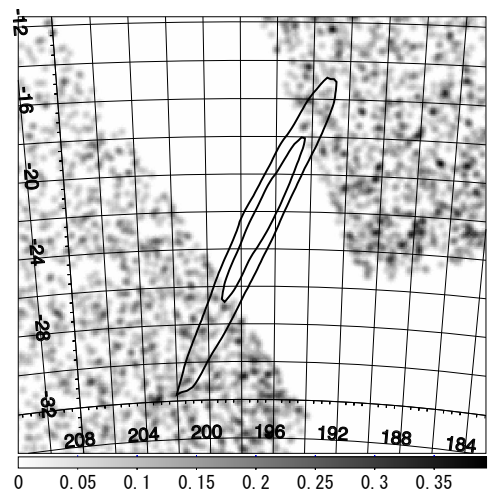}}
 \subfigure[one-day observation]{
  \includegraphics[width=6.5cm] {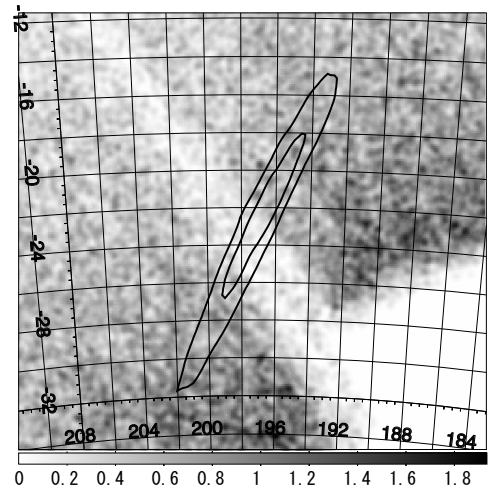}}
 \subfigure[ten-day observation]{
 \includegraphics[width=6.5cm] {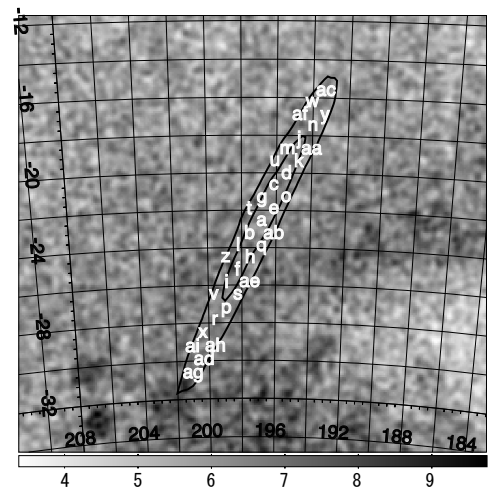}}
  \end{center}
\caption{GSC X-ray images at 2--20 keV with the 90\% and 50\% probability contours of GW170817 (thick lines) in three intervals of one orbit (from T0+0 to T0+5520 s), one day, and ten days. The images are X-ray photon-count images with a resolution of 0.1 deg and smoothed. The ten-day image (c) also shows the positions using the upper-limit calculation, of which alphabets correspond to Table \ref{tbl:ul}.}
\label{fig:images}
\end{figure}

In the MAXI observation of GW170817, 
the high voltage of GSC was off at the GW trigger time of August 17, 2017 12:41:04 UTC.
It was turned on 173 s after the trigger time.
GSC covered 62\% of the 90\% probability source region during the first orbit (92 min) after the GW trigger.
Figure \ref{fig:images} shows X-ray images of the GSC observation in (a) the first orbit, (b) one day, and (c) ten days
with the 90\% and 50\% probability contours of localization by the LALInference v2 map of GW170817.
The MAXI/GSC real-time transient monitor and alert system (nova-alert system) is continuously operating \citep{2016PASJ...68S...1N}.
The nova-alert system detects transient events with fluxes $>$80 mCrab in a one-orbit scan 
and sends an alert to the world in less than 30 s after the onboard detection of the transient.
In the probability region of GW170817, no significant X-ray transient was detected by the system. \\
We calculated the upper limits of the source detection with 3 $\sigma$ significance in the probability region 
according to the following procedure.
We selected 33 points in the probability region for upper-limit calculation using the HEALPix library \citep{2005ApJ...622..759G}, 
which were spherical surface pixels divided into 49152 pixels ($N_{side} = 64$).
Figure \ref{fig:images} (c) shows the 33 points with HEALPix numbers at $N_{side} = 64$.
We estimated the source photon counts $C_{\rm src}$ in each point for detection with 3 $\sigma$ significance from the background $C_{\rm bg}$.
Details of $C_{\rm src}$ estimation are described in appendix \ref{sec:ul}.
The effective exposure $EE$ (cm$^{2}$ s) was calculated by time integration of the geometrically corrected effective area in the FOV
since the effective area of GSC toward a source continuously changes during a scan \citep{2011PASJ...63S.623M}.
The photon-flux upper limit $f_{\rm UL}$ with $N \sigma$ significance was calculated as $\displaystyle f_{\rm UL} = \frac{C_{\rm src}(N)}{EE}$.
Table \ref{tbl:ul} lists the 3 $\sigma$ upper limits of the 33 points for one-orbit, one-day, and ten-day observation within the 90\% probability region of GW170817.
The average upper limit in the 33 points was 53/26 mCrab (one-orbit observation/one-day observation).

 \begin{longtable}{ccc crrr crrr crrr}
  \caption{GSC X-ray flux upper limits in the 2--10 keV band at the 90\% probability region of GW170817}
    \label{tbl:ul}
 \endhead
  \hline
& & & \multicolumn{4}{c}{one-orbit observation} &  \multicolumn{4}{c}{one-day observation} & \multicolumn{4}{c}{ten-day observation}   \\
  \cline{4-7} \cline{8-11} \cline{12-16} 
  ID\footnotemark[$*$] & Probability & RA, Dec\footnotemark[$\dag$]  
  & cam\footnotemark[$\ddag$] & $C_{\rm bg}$\footnotemark[$\S$] & $EE$\footnotemark[$\P$] & $f_{\rm U.L}$\footnotemark[$\Vert$]
    & cam & $C_{\rm bg}$ & $EE$ & $f_{\rm U.L}$ 
    & cam & $C_{\rm bg}$ & $EE$ & $f_{\rm U.L}$
    \\
    \hline
    \endhead
    \multicolumn{15}{l}{\footnotemark[$*$] point ID shown in Figure \ref{fig:images} (c)} \\
    \multicolumn{15}{l}{\footnotemark[$\dagger$] position of the point in J2000 coordinates} \\
    \multicolumn{15}{l}{\footnotemark[$\ddagger$] ID of the GSC camera} \\
    \multicolumn{15}{l}{\footnotemark[$\S$] observed counts in a PSF of GSC} \\
    \multicolumn{15}{l}{\footnotemark[$\P$] effective exposure (cm$^{2}$s)} \\
    \multicolumn{15}{l}{\footnotemark[$\Vert$] 3$\sigma$ upper limit in the 2--10 keV band (mCrab)} \\
    \endlastfoot
    \hline
a & 2.71$\times 10^{-4}$ & 196.88, -22.67 & - & - & - & - & 2,5 & 121 & 471 & 43 & 2,4,5 & 2467 & 20373 & 4 \\ 
b & 2.11$\times 10^{-4}$ & 197.58, -23.32 & - & - & - & - & 2,5 & 85 & 339 & 51 & 2,4,5 & 2488 & 20630 & 4 \\ 
c & 2.07$\times 10^{-4}$ & 196.17, -20.74 & - & - & - & - & 2 & 195 & 1008 & 25 & 2,4,5 & 2420 & 19988 & 4 \\ 
d & 1.99$\times 10^{-4}$ & 195.47, -20.11 & - & - & - & - & 2 & 213 & 1121 & 23 & 2,4,5 & 2503 & 19584 & 4 \\ 
e & 1.99$\times 10^{-4}$ & 196.17, -22.02 & - & - & - & - & 2,5 & 172 & 842 & 28 & 2,4,5 & 2499 & 20115 & 4 \\ 
f & 1.74$\times 10^{-4}$ & 198.28, -25.28 & - & - & - & - & 2,5 & 153 & 750 & 30 & 2,4,5 & 2694 & 21814 & 4 \\ 
g & 1.48$\times 10^{-4}$ & 196.88, -21.38 & - & - & - & - & 2,5 & 169 & 777 & 30 & 2,4,5 & 2425 & 20209 & 4 \\ 
h & 1.39$\times 10^{-4}$ & 197.58, -24.62 & - & - & - & - & 2,5 & 97 & 370 & 50 & 2,4,5 & 2573 & 21020 & 4 \\ 
i & 1.32$\times 10^{-4}$ & 198.98, -25.94 & 5 & 19 & 215 & 43 & 5 & 233 & 1309 & 21 & 4,5 & 2861 & 22576 & 4 \\ 
j & 1.22$\times 10^{-4}$ & 194.77, -18.21 & - & - & - & - & 2 & 254 & 1336 & 21 & 2,4,5 & 2611 & 19493 & 4 \\ 
k & 1.18$\times 10^{-4}$ & 194.77, -19.47 & 2 & 5 & 156 & 38 & 2 & 238 & 1266 & 22 & 2,4,5 & 2547 & 19379 & 4 \\ 
l & 1.16$\times 10^{-4}$ & 198.28, -23.97 & - & - & - & - & 2,5 & 95 & 422 & 43 & 2,4,5 & 2597 & 21059 & 4 \\ 
m & 1.12$\times 10^{-4}$ & 195.47, -18.84 & 2 & 3 & 153 & 34 & 2 & 232 & 1196 & 23 & 2,4,5 & 2577 & 19352 & 4 \\ 
n & 9.53$\times 10^{-5}$ & 194.06, -17.58 & 2 & 25 & 151 & 68 & 2 & 272 & 1452 & 20 & 2,4,5 & 2549 & 19604 & 4 \\ 
o & 9.40$\times 10^{-5}$ & 195.47, -21.38 & - & - & - & - & 2 & 193 & 1057 & 24 & 2,4,5 & 2445 & 19906 & 4 \\ 
p & 9.37$\times 10^{-5}$ & 198.98, -27.28 & 5 & 31 & 220 & 51 & 5 & 308 & 1681 & 18 & 4,5 & 2864 & 23004 & 4 \\ 
q & 9.33$\times 10^{-5}$ & 196.88, -23.97 & - & - & - & - & 2,5 & 78 & 337 & 50 & 2,4,5 & 2485 & 20642 & 4 \\ 
r & 7.94$\times 10^{-5}$ & 199.69, -27.95 & 5 & 44 & 219 & 60 & 5 & 325 & 1855 & 17 & 4,5 & 2904 & 23286 & 4 \\ 
s & 7.40$\times 10^{-5}$ & 198.28, -26.61 & 5 & 22 & 221 & 44 & 5 & 232 & 1243 & 22 & 4,5 & 2777 & 22561 & 4 \\ 
t & 7.24$\times 10^{-5}$ & 197.58, -22.02 & - & - & - & - & 2,5 & 119 & 444 & 45 & 2,4,5 & 2411 & 20434 & 4 \\ 
u & 6.48$\times 10^{-5}$ & 196.17, -19.47 & - & - & - & - & 2 & 209 & 1070 & 24 & 2,4,5 & 2488 & 19656 & 4 \\ 
v & 6.10$\times 10^{-5}$ & 199.69, -26.61 & 5 & 29 & 214 & 51 & 5 & 292 & 1668 & 18 & 4,5 & 2922 & 22948 & 4 \\ 
w & 5.52$\times 10^{-5}$ & 194.06, -16.33 & 2 & 29 & 147 & 75 & 2 & 248 & 1463 & 19 & 2,4,5 & 2507 & 19348 & 4 \\ 
x & 5.49$\times 10^{-5}$ & 200.39, -28.63 & 5 & 42 & 219 & 59 & 5 & 320 & 1937 & 16 & 4,5 & 2894 & 23507 & 3 \\ 
y & 4.74$\times 10^{-5}$ & 193.36, -16.96 & 2 & 27 & 149 & 72 & 2 & 243 & 1497 & 18 & 2,4,5 & 2480 & 19130 & 4 \\ 
z & 4.57$\times 10^{-5}$ & 198.98, -24.62 & - & - & - & - & 2,5 & 153 & 851 & 26 & 2,4,5 & 2690 & 21916 & 4 \\ 
aa & 4.48$\times 10^{-5}$ & 194.06, -18.84 & 2 & 14 & 155 & 53 & 2 & 270 & 1411 & 21 & 2,4,5 & 2540 & 19543 & 4 \\ 
ab & 4.36$\times 10^{-5}$ & 196.17, -23.32 & - & - & - & - & 2,5 & 124 & 508 & 40 & 2,4,5 & 2446 & 20314 & 4 \\ 
ac & 4.33$\times 10^{-5}$ & 193.36, -15.71 & 2 & 31 & 147 & 77 & 2 & 236 & 1468 & 19 & 2,4,5 & 2389 & 18850 & 4 \\ 
ad & 3.79$\times 10^{-5}$ & 200.39, -30.00 & 5 & 44 & 227 & 58 & 5 & 350 & 2197 & 15 & 4,5 & 3022 & 24019 & 3 \\ 
ae & 3.76$\times 10^{-5}$ & 197.58, -25.94 & - & - & - & -& 2,5 & 152 & 652 & 34 & 2,4,5 & 2699 & 21748 & 4 \\ 
af & 3.74$\times 10^{-5}$ & 194.77, -16.96 & 2 & 19 & 150 & 62 & 2 & 257 & 1384 & 20 & 2,4,5 & 2633 & 19668 & 4 \\ 
ag & 3.47$\times 10^{-5}$ & 201.09, -30.69 & 5 & 46 & 224 & 59 & 5 & 373 & 2341 & 14 & 4,5 & 3314 & 24219 & 4 \\ 
ah & 2.98$\times 10^{-5}$ & 199.69, -29.31 & 5 & 41 & 225 & 56 & 5 & 328 & 1994 & 16 & 4,5 & 2962 & 23642 & 3 \\ 
ai & 2.95$\times 10^{-5}$ & 201.09, -29.31 & 5 & 36 & 220 & 54 & 5 & 343 & 2125 & 15 & 4,5 & 2985 & 23856 & 3 \\ 
    \hline
\end{longtable}

\subsection{Observation of the electromagnetic counterpart SSS17a}

The optical counterpart SSS17a was detected at (RA,Dec) = (\timeform{197D.45}, \timeform{-23D.38}) in the GW170817 localization area \citep{Coultereaap9811}.
The observational window of GSC at the position started at 20 s after the GW trigger,
but the high voltage of GSC was unfortunately off at the time because the ISS was entering a high-particle-background region with a latitude higher than 40 deg.
The high voltage turned on 20 s after the counterpart was out of the FOV.
The left panel of Figure \ref{fig:fov_ocp} (a) shows the GSC image at the first scan since the GW trigger 
and the PSF of GSC to the direction of SSS17a.
The SSS17a position was not observed until the third scan since the GW trigger because of the high-voltage-off operation at the region. \\
The first observation of the position was  performed 16797 s (4.6 h) after the GW trigger,
and it was the earliest X-ray observation of the counterpart \citep{2017ApJ...848L..12A}.
The left panels of Figure \ref{fig:fov_ocp} (b) -- (g) show the images of the scans around the position of SSS17a 
when the FOV of GSC pointed to SSS17a and the high voltage was on.
Until the scan at 49021 s, the observation did not cover the position with the full PSF because of the high-voltage-off operation.\\
Previously, we had conservatively reported on the upper limit of the scan in the full-PSF observation by the GCN \citep{2017GCN.21555}.
In the present study, we estimated the upper limits of the observations in case of a partial-PSF-coverage observation
in which the total effective exposure in each scan was larger than 1 cm$^{2}$ s.
We used the advanced good time interval (GTI) region for this analysis, in contrast to the standard analysis of GSC.
In the standard pipeline process, the GTI starts 15 s after 
the value of high-voltage house-keeping data reaches the set value of 1550 V.
Since the house-keeping data are measured at 10-s intervals,
the actual high voltage is already stable to the set value before the time of the house-keeping data.
To maximize the number of photon counts from SSS17a,
we adopted the new GTI, which was selected based on the criterion that the veto count rates were stable at approximately 300 counts/s.
The right panels of Figure \ref{fig:fov_ocp} show the time profile of the effective area of SSS17a and the veto count rates.
By using the new GTI, the photon counts and effective area in the PSF were greater than those in the case of standard analysis, 
but the photon counts of the partial-PSF-coverage region were not statistically significant to estimate the upper limit.
We adopted the photon counts of the area near SSS17a as the background counts.
The number of photon counts $C_{\rm bg}$ was scaled by the partial coverage factor of the $EE$ of SSS17a.
Table \ref{tbl:ul_ocp} lists the 3 $\sigma$ upper limits of energy flux of each scan of the position of SSS17a in the 2--10 keV band.

\begin{figure*}
 \begin{center} 
 \subfigure[at $T_{\rm GW}$ + 77 s]{
 \includegraphics[width=6cm] {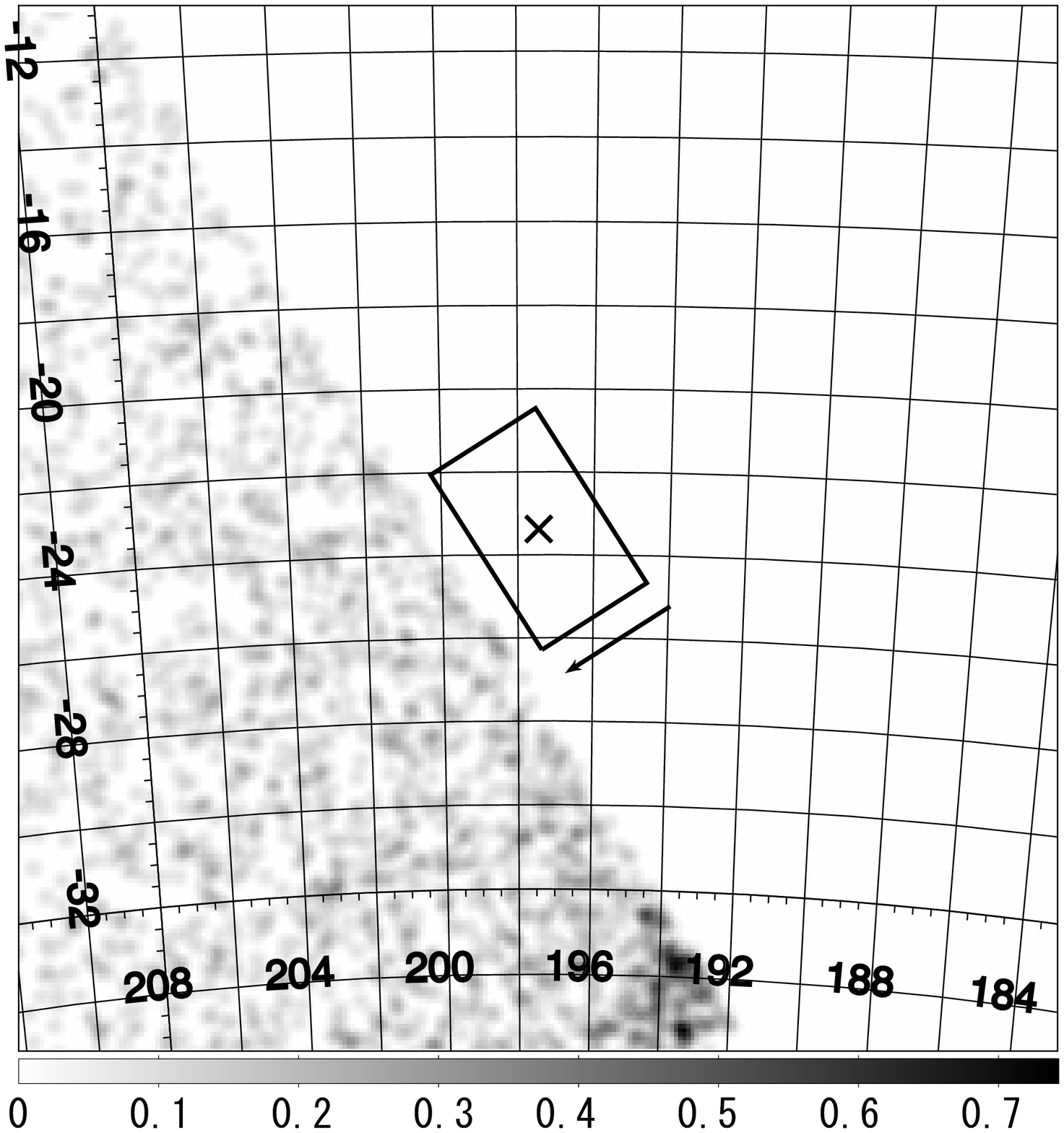}
 \includegraphics[width=8cm] {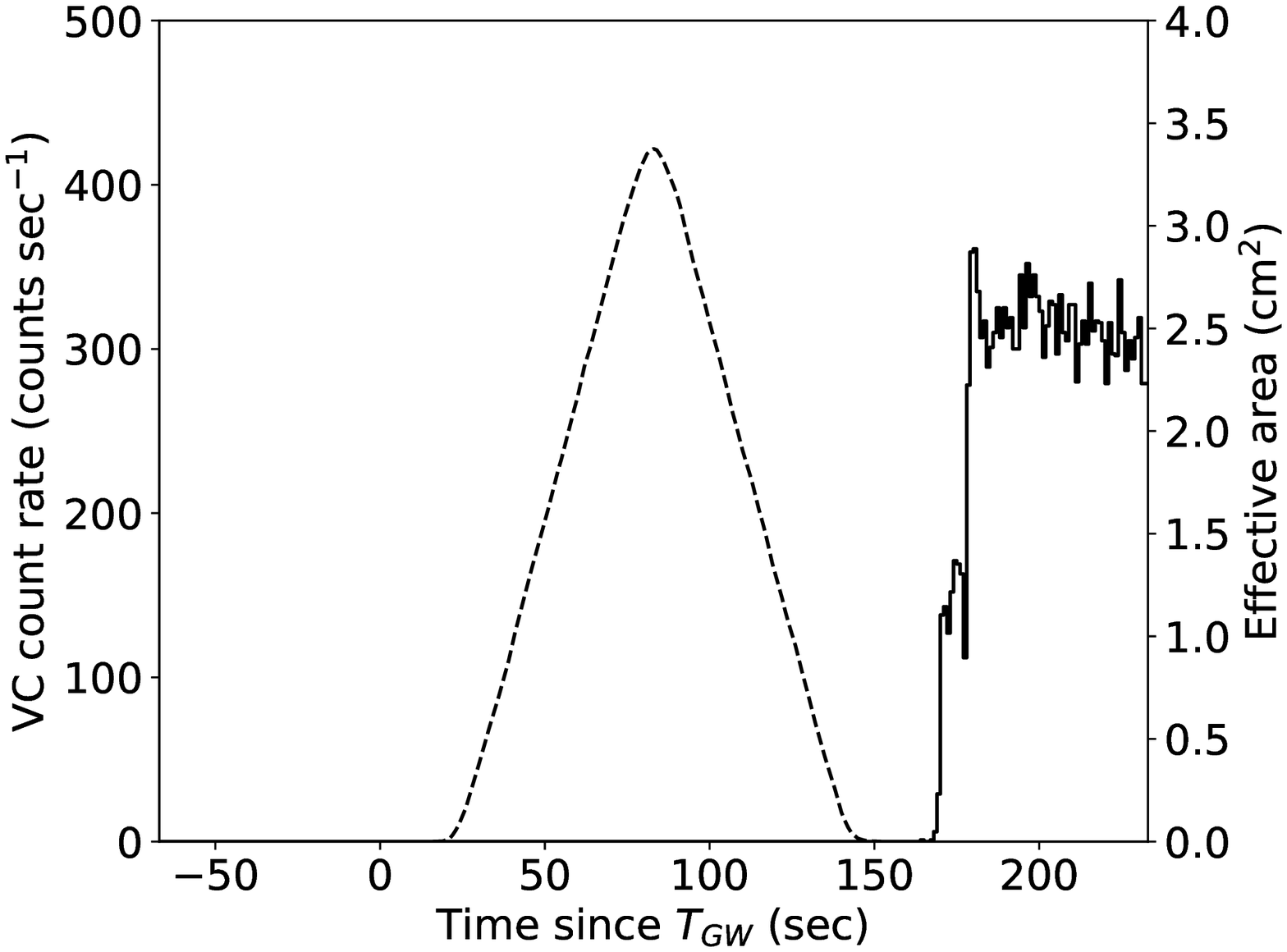}}
 \subfigure[at $T_{\rm GW}$ + 16797 s]{
 \includegraphics[width=6cm] {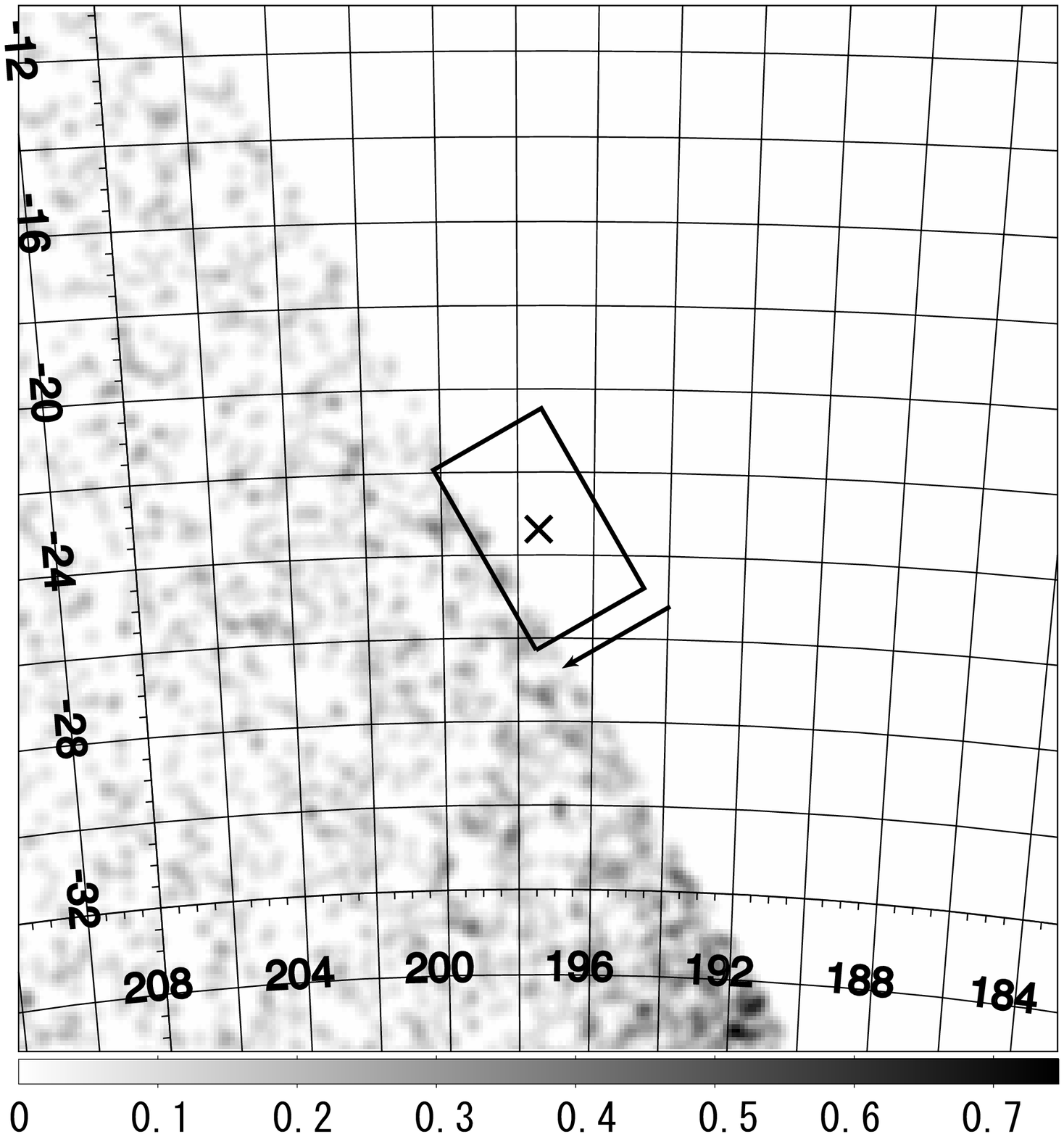}
 \includegraphics[width=8cm] {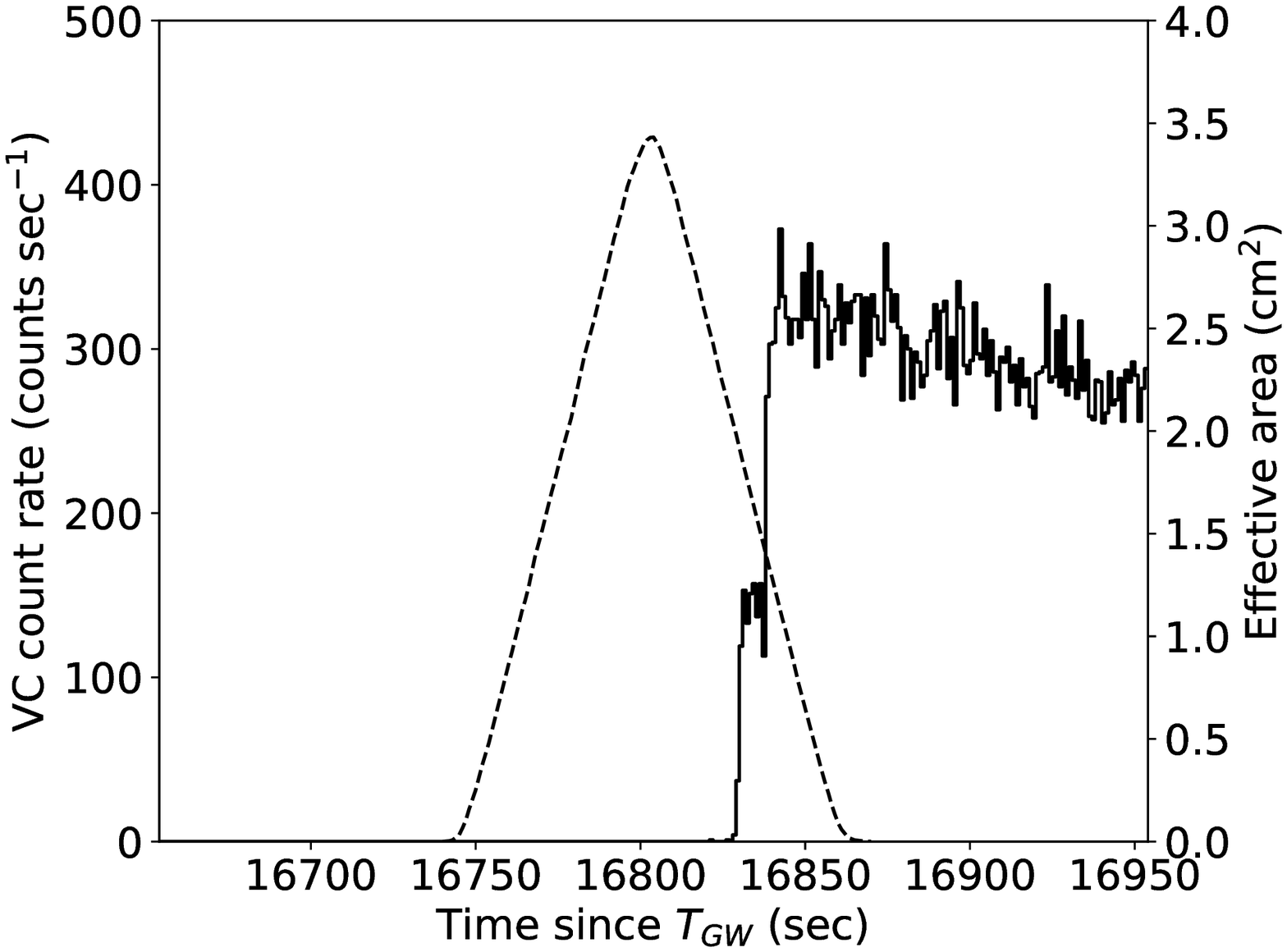}}
 \subfigure[at $T_{\rm GW}$ + 22344 s]{ 
 \includegraphics[width=6cm] {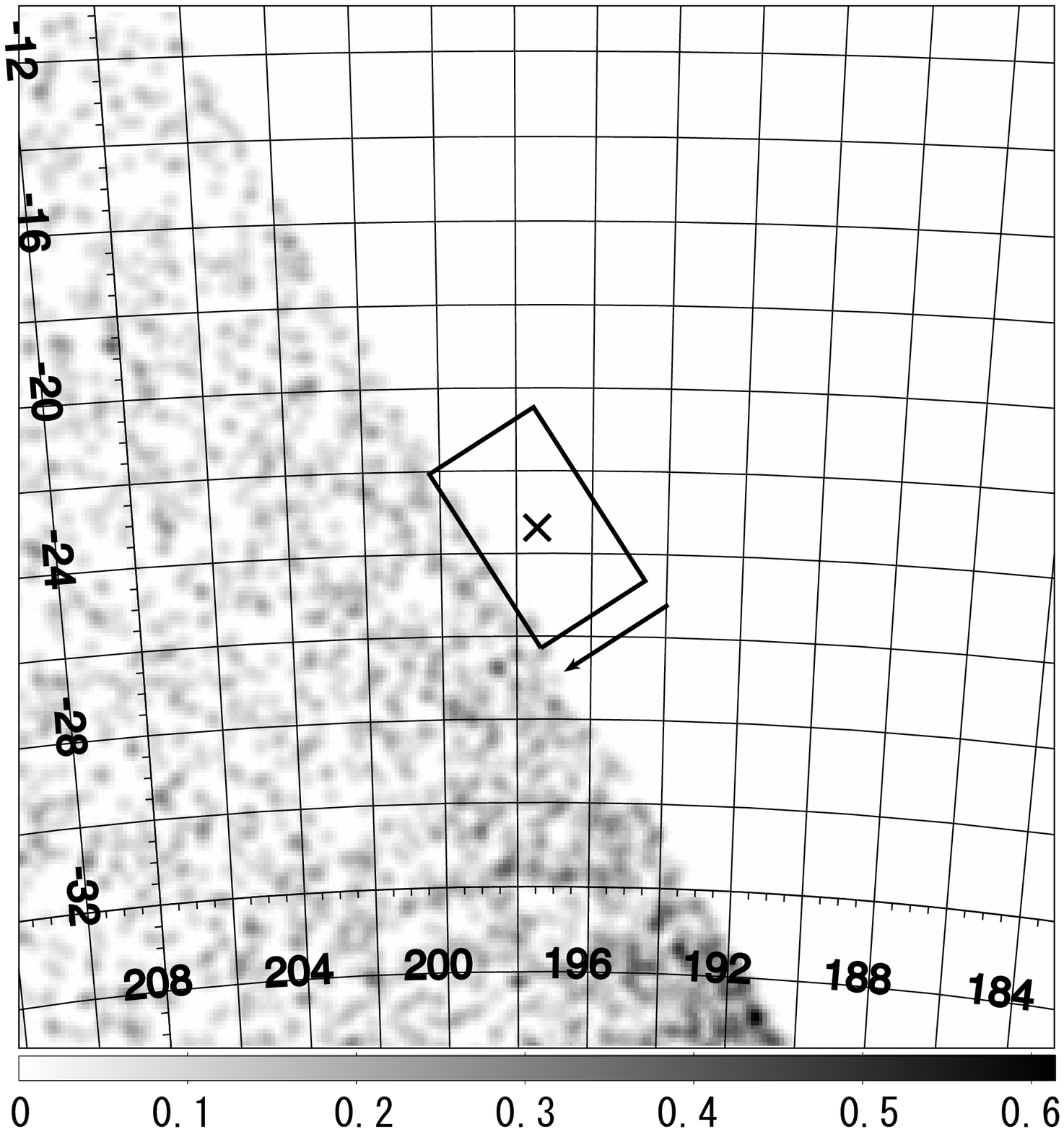}
 \includegraphics[width=8cm] {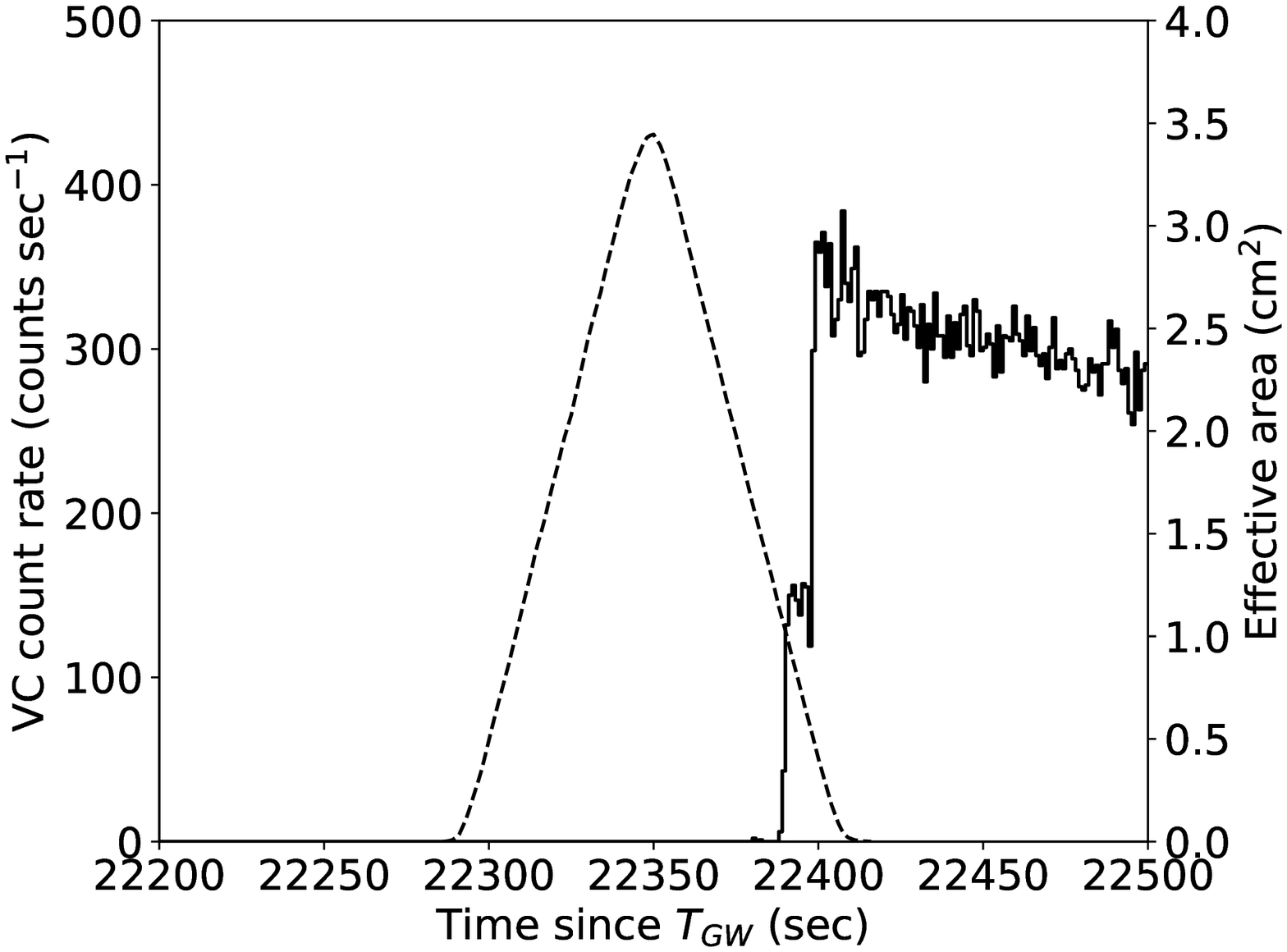}}
 \end{center}
\caption{Left: GSC X-ray images at 2--20 keV around the electromagnetic counterpart of GW170817. The point marked by ''x'' is the position of the optical counterpart SSS17a, the square region is the PSF of GSC at the position, and the arrow shows the scan direction of GSC.} Right: time profiles of the effective area of GSC to the counterpart (dashed line) and veto count rate of GSC (continuous line).
\label{fig:fov_ocp}
\end{figure*}

\addtocounter{figure}{-1}

\begin{figure*}
 \begin{center}
 \addtocounter{subfigure}{3}
 \subfigure[at $T_{\rm GW}$ + 43465 s]{
 \includegraphics[width=6cm] {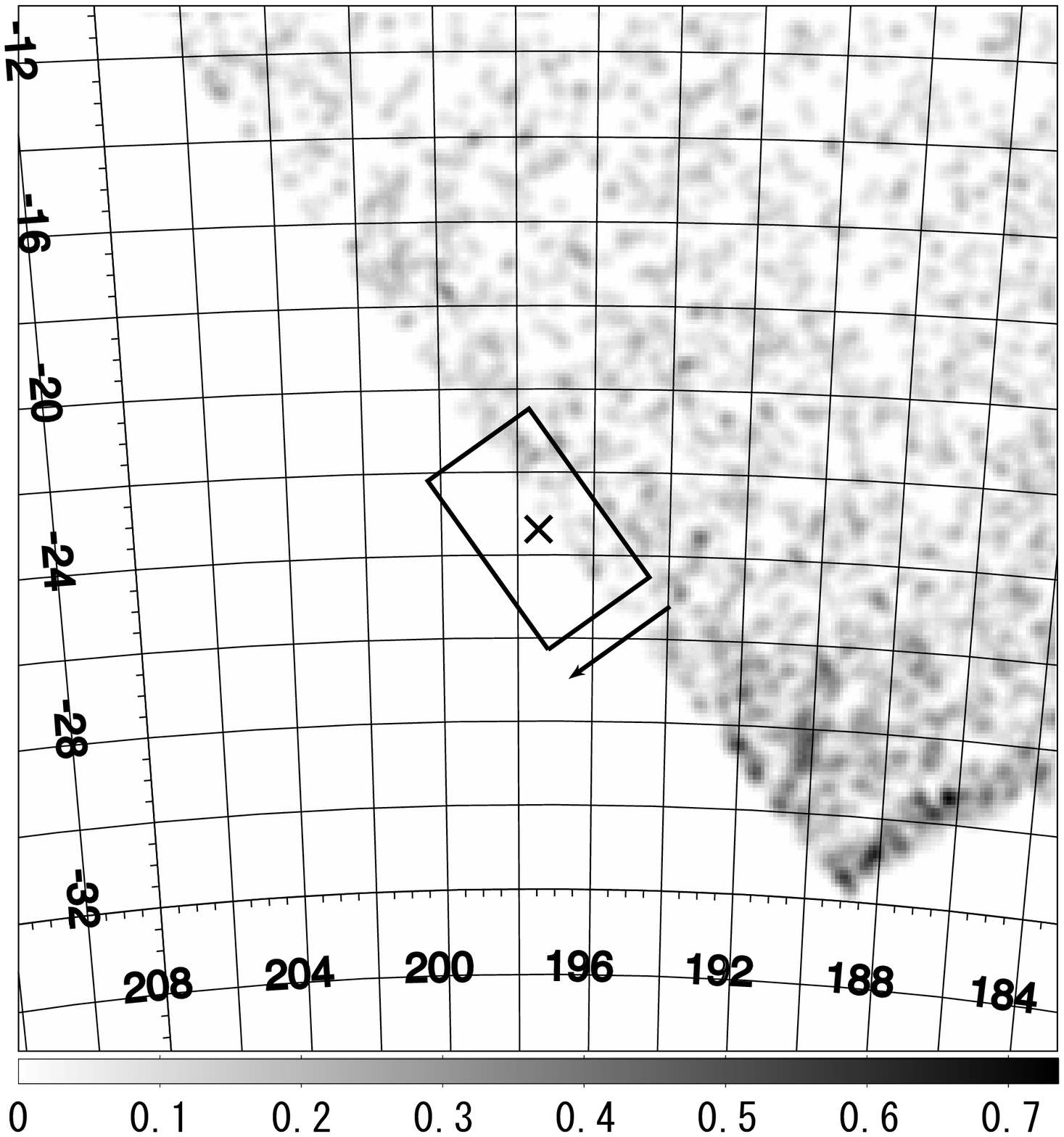}
 \includegraphics[width=8cm] {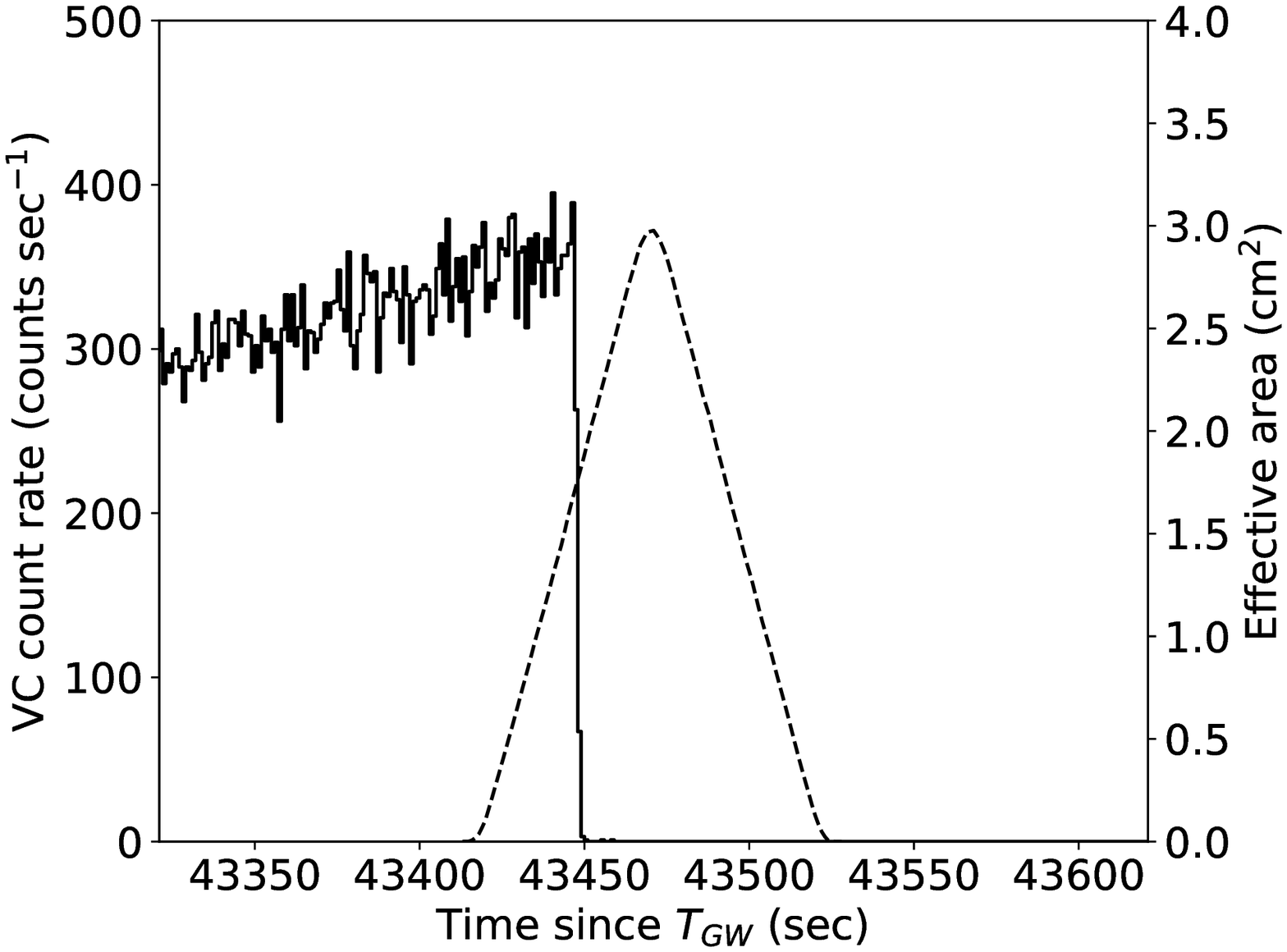}}
 \subfigure[at $T_{\rm GW}$ + 49021 s]{
 \includegraphics[width=6cm] {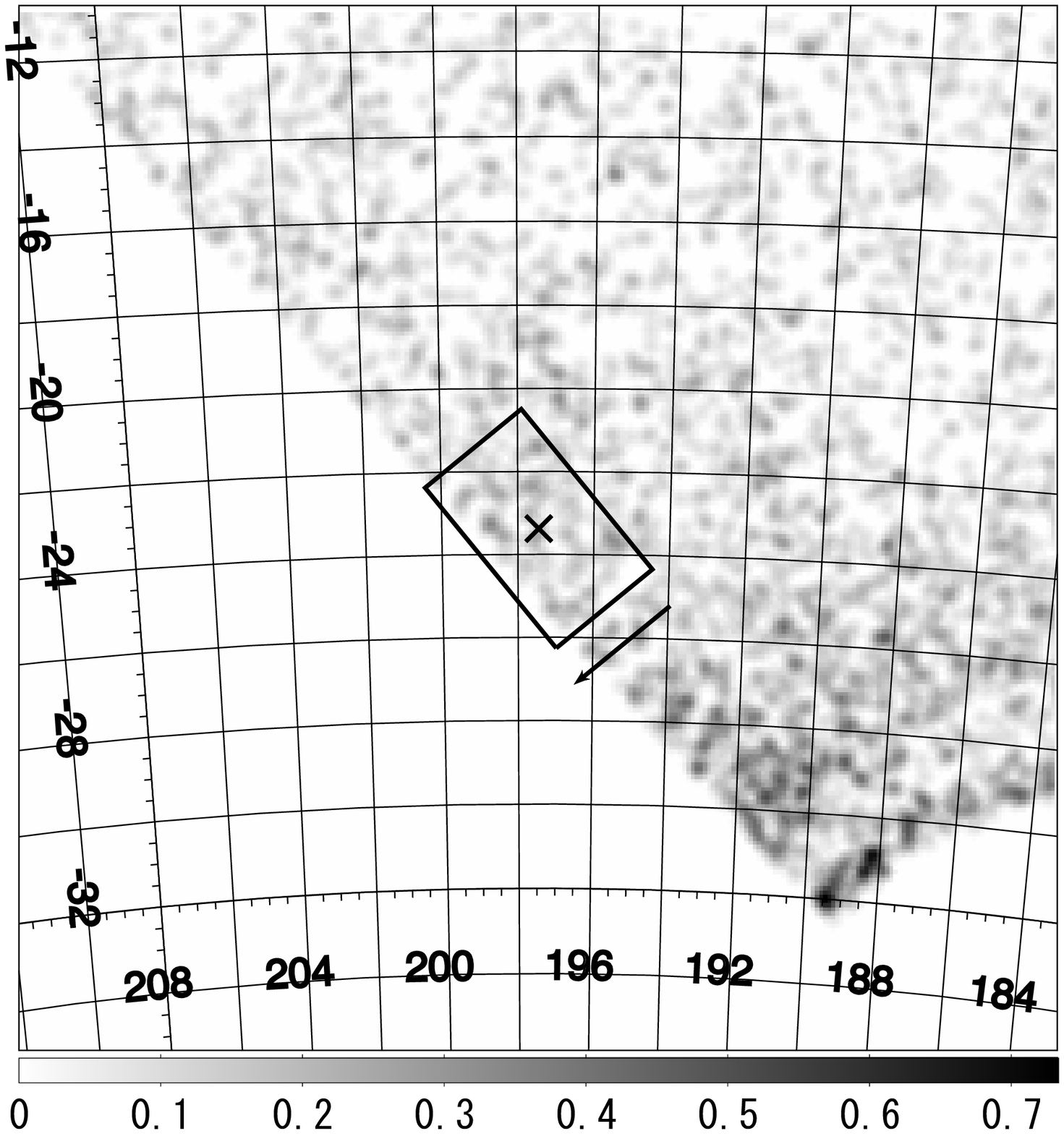}
 \includegraphics[width=8cm] {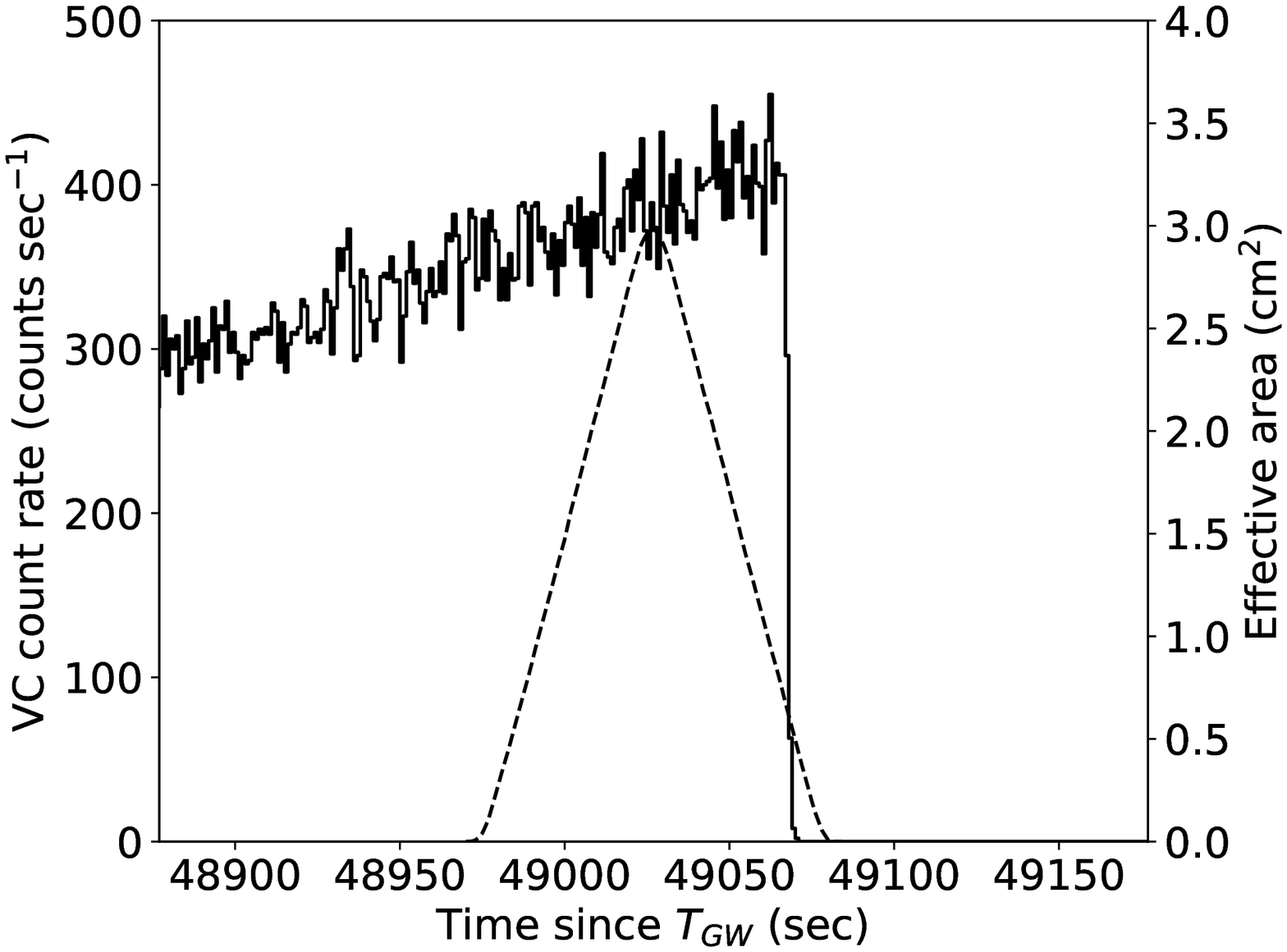}}
 \subfigure[at $T_{\rm GW}$ + 61319 s]{
 \includegraphics[width=6cm] {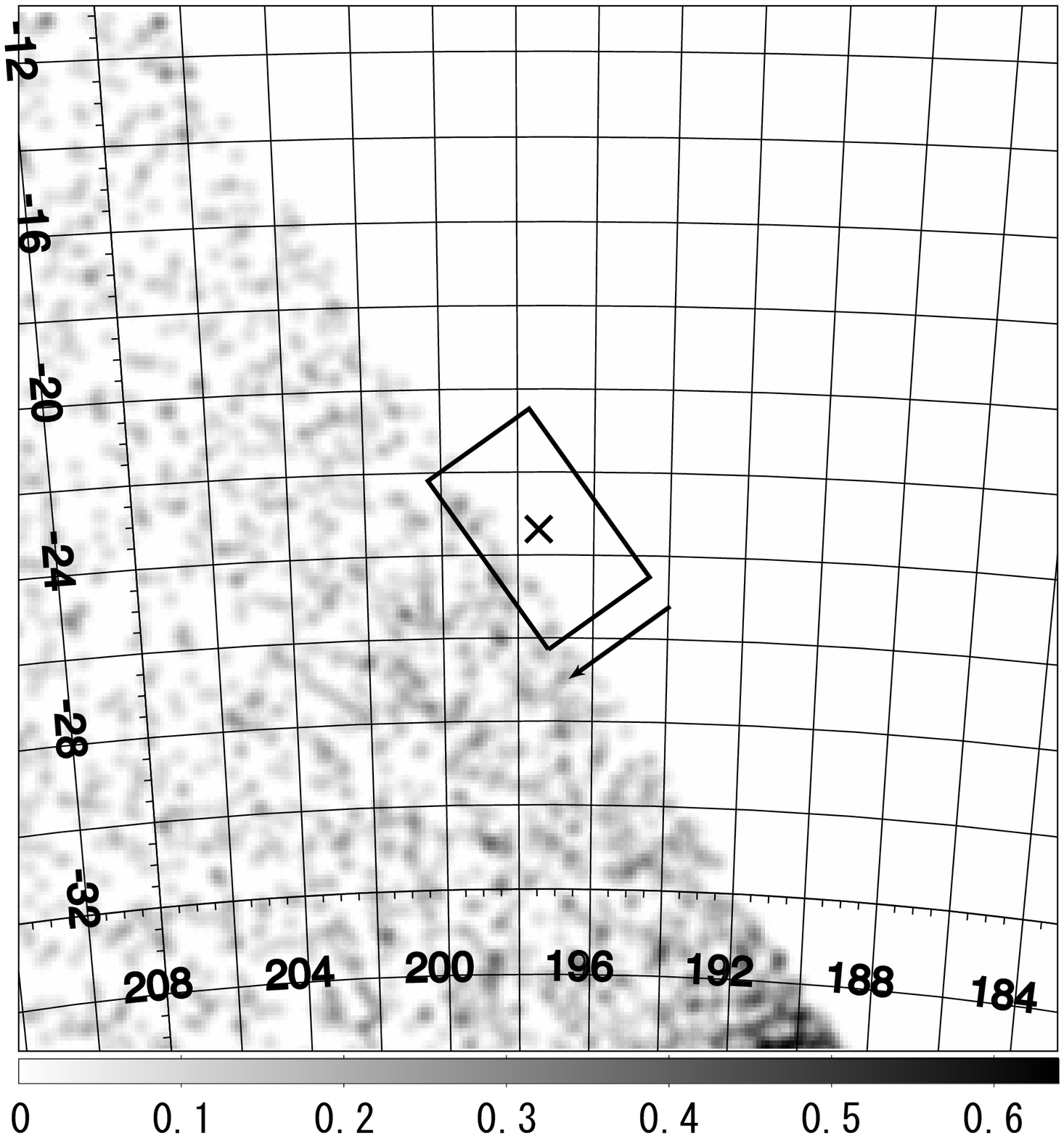}
 \includegraphics[width=8cm] {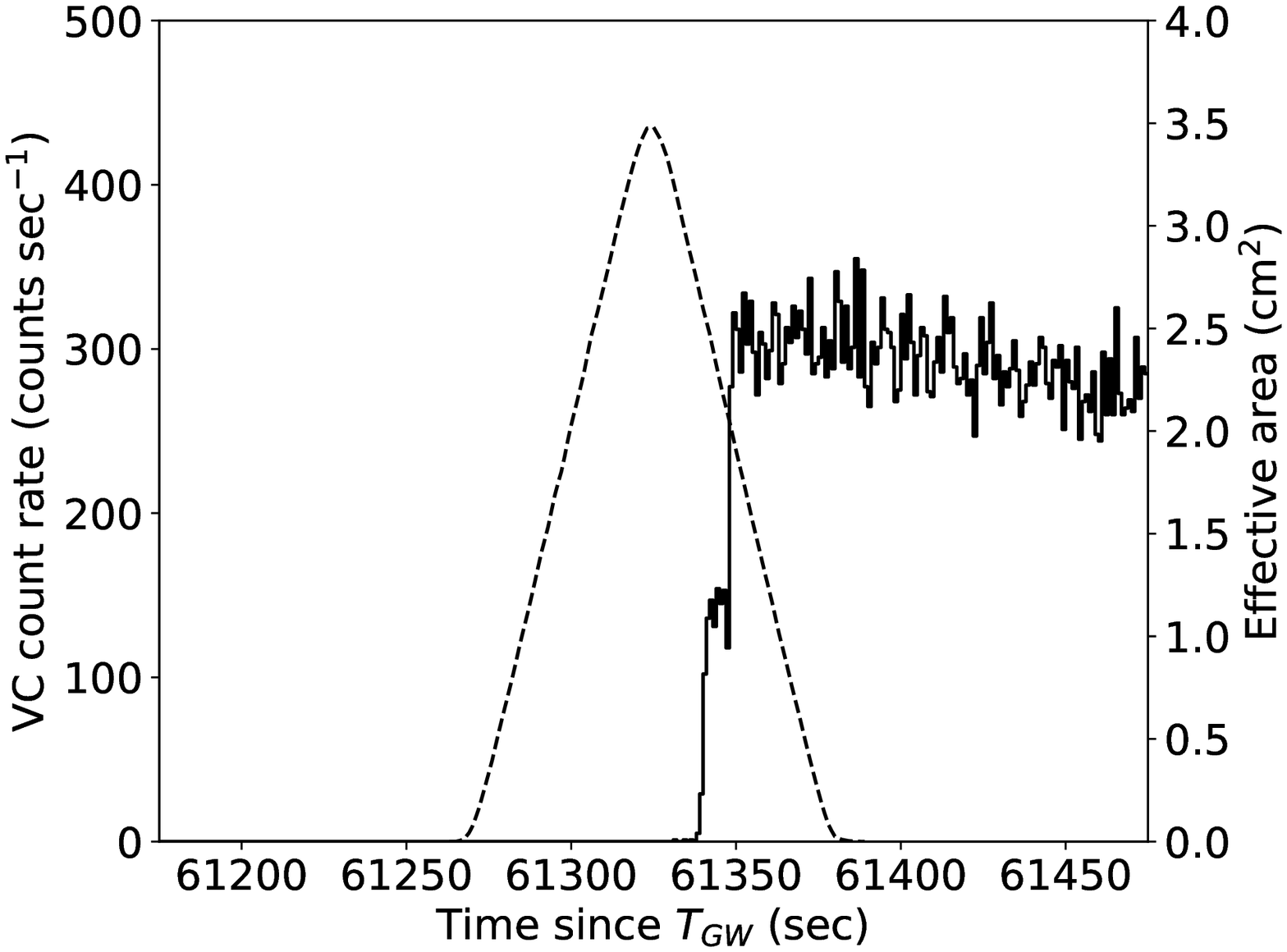}}
 \end{center}
\caption{(continued)}
\label{fig:fov_ocp}
\end{figure*}

\addtocounter{figure}{-1}

\begin{figure*}
 \begin{center}
 \addtocounter{subfigure}{6}
 \subfigure[at $T_{\rm GW}$ + 66886 s]{
 \includegraphics[width=6cm] {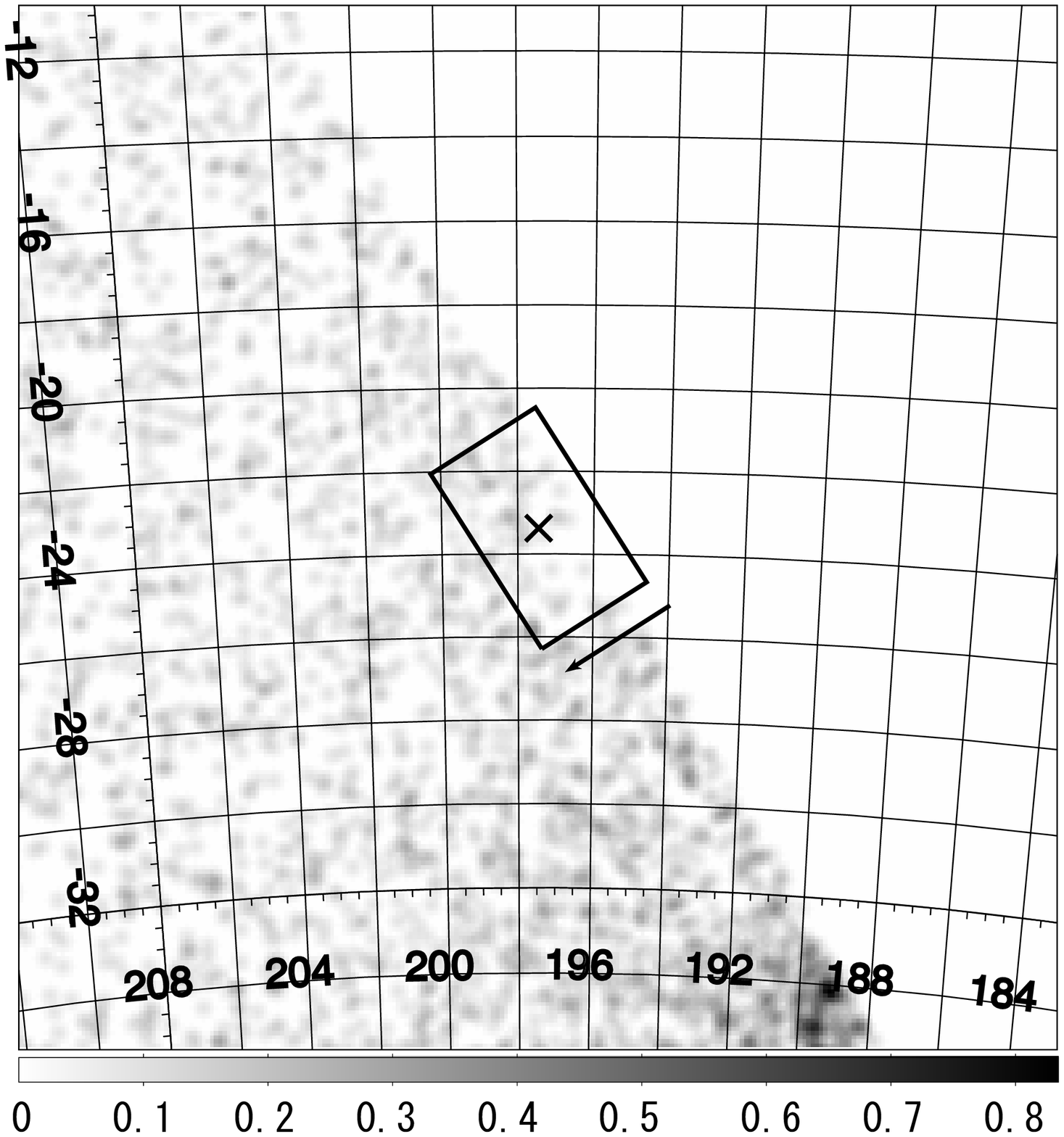}
 \includegraphics[width=8cm] {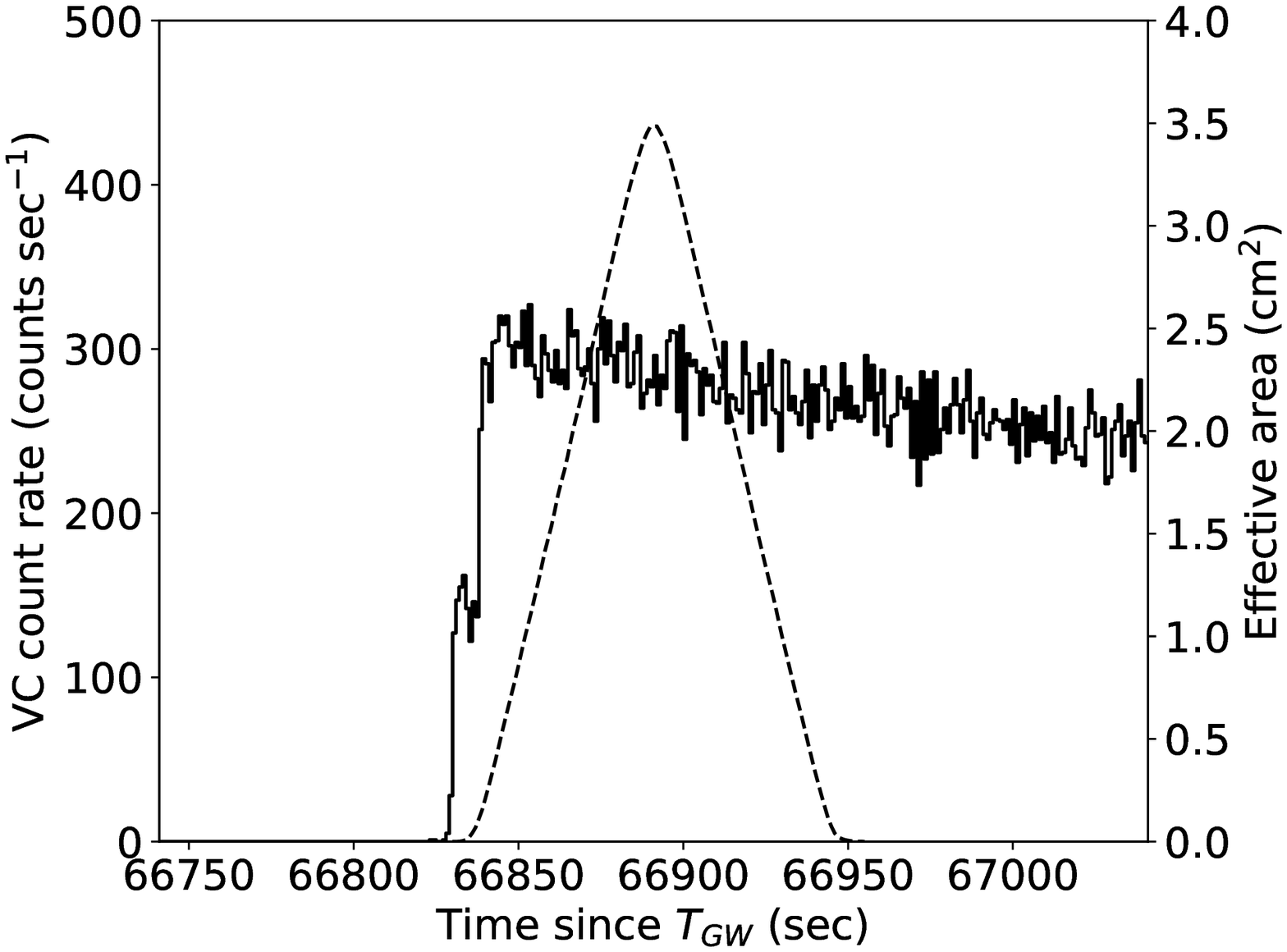}}
 \end{center}
\caption{(continued)}
\label{fig:fov_ocp}
\end{figure*}

\begin{table*}[htp]
\caption{GSC X-ray flux upper limit in the 2--10 keV band at the position of the electromagnetic counterpart}  
\begin{center}
\begin{tabular}{cccccc}
\hline \hline
Time since trigger (s) & cam\footnotemark[$*$] & $C_{\rm bg}$\footnotemark[$\dag$] & $EE$\footnotemark[$\ddag$] & $f_{\rm U.L} $(erg cm$^{-2}$ s$^{-1}$)\footnotemark[$\S$] & $L_{\rm U.L}$ (erg s$^{-1}$)\footnotemark[$\P$] \\
\hline
16797 & 5 & 73 & 12 & 8.60$\times 10^{-9}$ & 1.65$\times 10^{45}$ \\
22344 & 5 & 76 & 1 & 7.70$\times 10^{-8}$ & 1.47$\times 10^{46}$ \\
43465 & 2 & 72 & 44 & 4.20$\times 10^{-9}$ & 8.00$\times 10^{44}$ \\
49021 & 2 & 73 & 161 & 2.17$\times 10^{-9}$ & 4.20$\times 10^{44}$ \\
61319 & 5 & 75 & 26 & 4.91$\times 10^{-9}$ & 9.41$\times 10^{44}$ \\
66886 & 5 & 62 & 198 & 1.52$\times 10^{-9}$ & 2.91$\times 10^{44}$ \\
\hline
\end{tabular}
\end{center}
\footnotemark[$*$] ID of the GSC camera \\
\footnotemark[$\dag$] observed counts in an FOV \\
\footnotemark[$\ddag$] effective exposure (cm$^2 $s) \\
\footnotemark[$\S$] 3$\sigma$ upper limit of X-ray flux in the 2--10 keV band\\
\footnotemark[$\P$] 3$\sigma$ upper limit of luminosity in the 2--10 keV band at a distance of 40 Mpc\\
\label{tbl:ul_ocp}
\end{table*}%

\section{Discussion and Conclusion}

\begin{figure}[htp]
 \begin{center}
 \includegraphics[width=8cm] {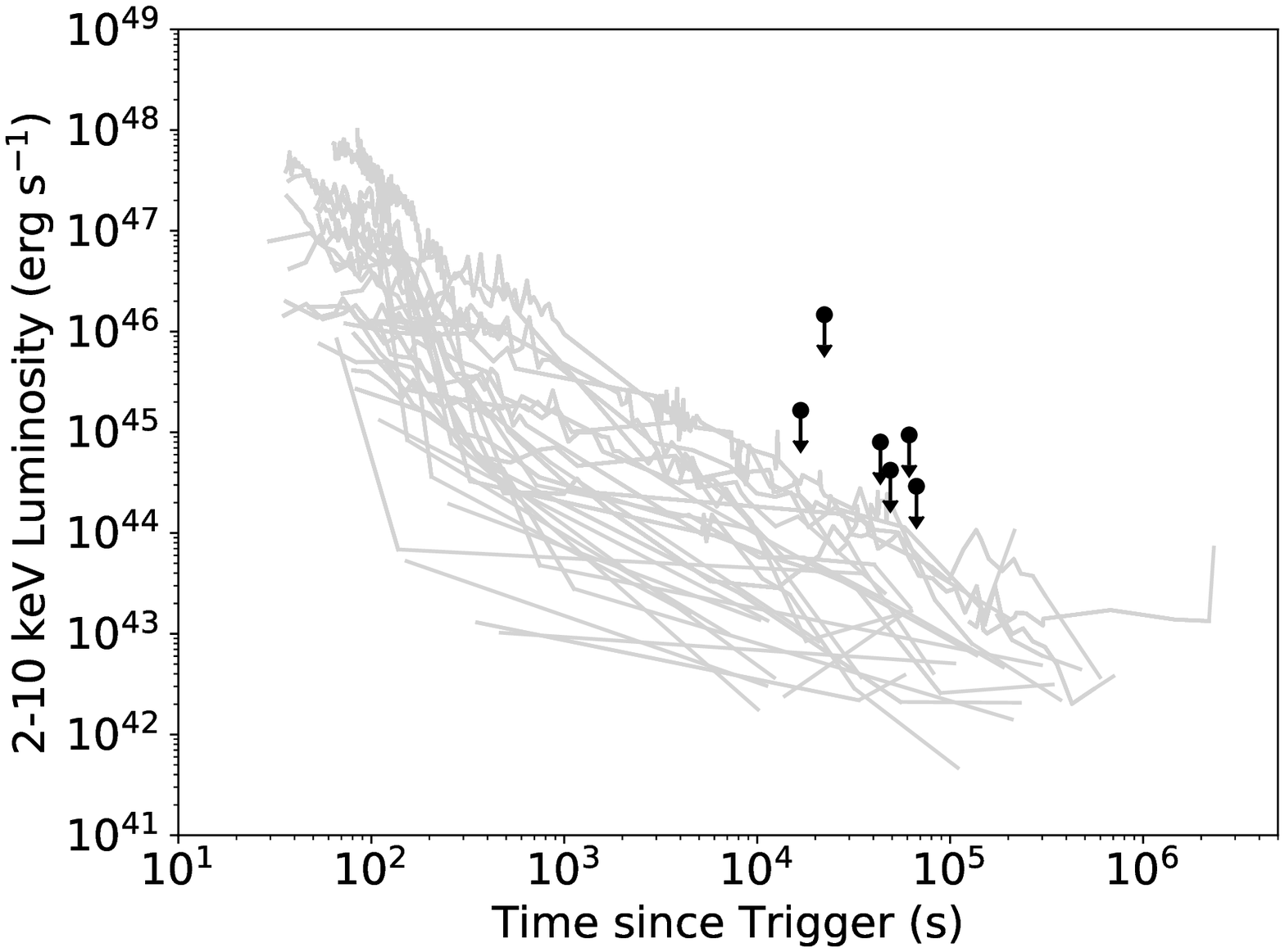}
 \end{center}
\caption{Upper limits of isotropic luminosity of GSC observations for SSS17a in comparison with canonical short GRB afterglows. 
The black points show the upper limits of GSC observation in one scan.
The gray lines are the lightcurves of canonical short on-axis GRB afterglows. }
\label{fig:sgrblc_lumi}
\end{figure}

\begin{figure}[htp]
 \begin{center}
 \includegraphics[width=8cm] {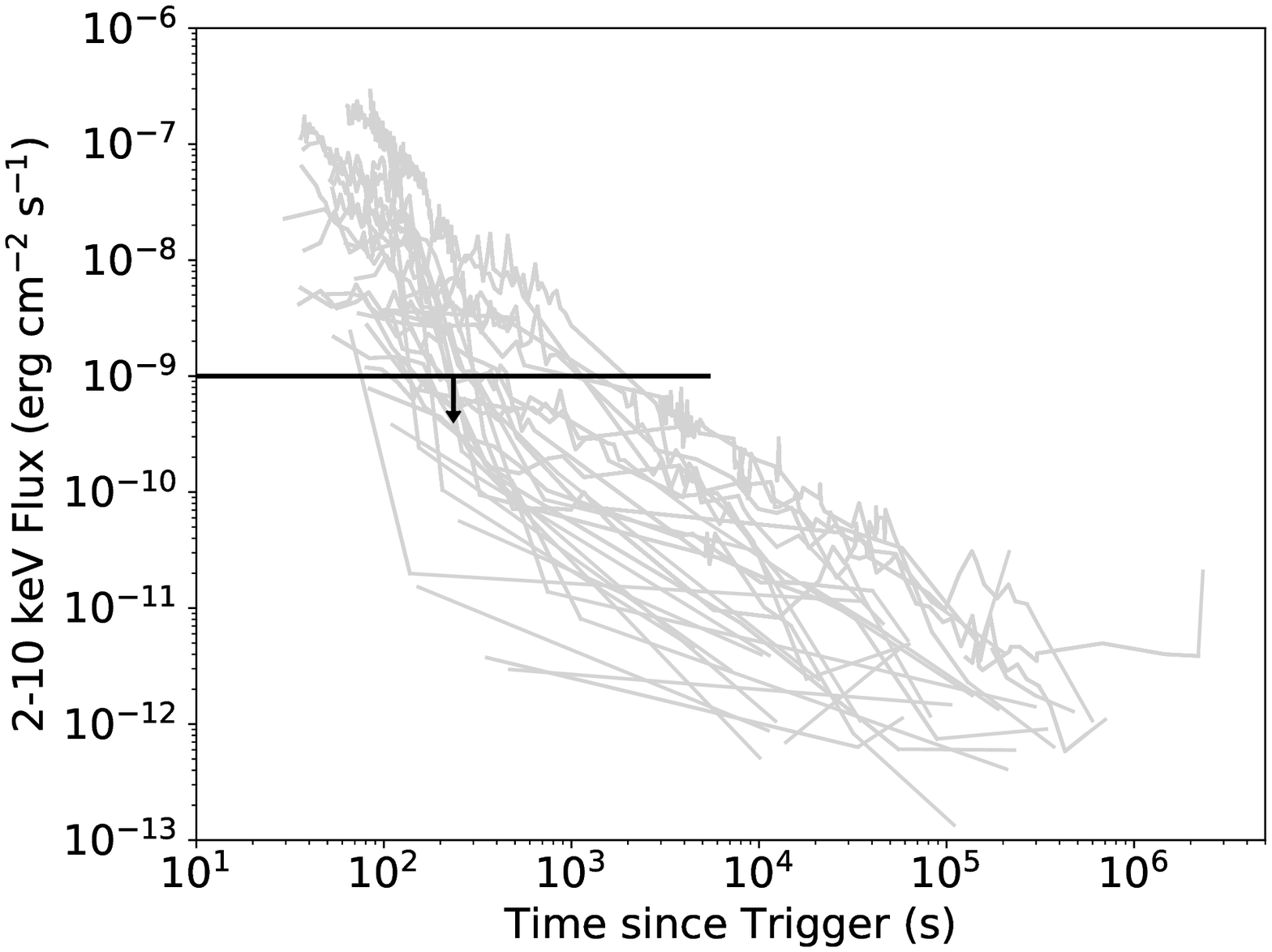}
 \end{center}
\caption{Typical flux upper limit of GSC observations in comparison with canonical short GRB afterglows.
The black line shows the typical upper limit with 3 $\sigma$ significance of GSC observation with full FOV in one scan.
The gray lines are the lightcurves of canonical short on-axis GRB afterglows, assuming a distance of 170 Mpc. }
\label{fig:sgrblc_flux}
\end{figure}

GRB170817A was detected by Fermi/GBM and INTEGRAL/SPI-ACS \citep{2017ApJ...848L..14G, 2017ApJ...848L..15S}
1.7 s after the GW170817 trigger \citep{2016PhRvL.116f1102A}.
The position of SSS17a was within the localization region of GRB170817A, 
implying that GRB170817A is a short GRB associated with a BNS merger.
We compared the upper limits of X-ray luminosity of GW170817 observed by the GSC with the luminosities of the short GRB afterglows. 
We used the data of Swift/XRT from the lightcurve repository \citep{2007A&A...469..379E}.
We calculated the upper limits of isotropic luminosity of GW170817, as listed in Table \ref{tbl:ul_ocp},  
as well as the lightcurves of isotropic luminosity of the canonical short GRBs by using the redshift listed by \citet{2015ApJ...815..102F}.
Figure \ref{fig:sgrblc_lumi} shows the time profile of the upper limits of X-ray luminosity of GW170817 observed by GSC 
and the X-ray afterglow luminosities of the canonical short GRBs.
The luminosity upper limits of the early X-ray observation of GW170817 by GSC were above the isotropic luminosities of the short GRB afterglows.\\
The X-ray upper limits observed by the GSC did not constrain the emission model 
of early X-ray afterglow of GW170817.
However, in case of the GSC observation of GW170817, 
the observational window unfortunately occurred during high-voltage-off operation, 
which occurs for approximately 15\% of the whole sky in one-orbit observation \citep{2011PASJ...63S.635S}.
In the future observation of the GW electromagnetic (EM) counterpart, 
we can expect
that GSC will observe the counterpart in the time of high-voltage-on operation with a probability of 85\%.
GSC observation in one scan typically has a sensitivity of $10^{-9}$ erg cm$^{-2}$ s$^{-1}$ in the 2--10 keV band.
Since MAXI (ISS) completes one orbit in 5520 s, the GSC can observe the counterpart with an exposure of $\sim$150 s within the 5520 s.
LIGO/Virgo will be upgraded for the next observing run 
(O3: one year run from fall 2018) to double the horizon distance up to 170 Mpc \citep{2016LRR....19....1A}.
Figure \ref{fig:sgrblc_flux} shows the typical sensitivity of GSC with 3 $\sigma$ significance in one scan as well as X-ray fluxes of the afterglows of the canonical short GRBs, assuming a distance of 170 Mpc.
If the GSC observes the EM counterpart in the first scan from the trigger,
the sensitivity would be sufficiently below the X-ray fluxes of the afterglow of the canonical short GRB. \\
It was reported that the characteristics of the X-ray afterglow of GW170817 are different from those of canonical short GRBs.
The late X-ray afterglow of GW170817 was detected in the Chandra observation \citep{2017ApJ...848L..20M, 2017Natur.551...71T, 2041-8205-848-2-L25}.
The flux of the afterglow 15 days after the GW trigger was 100 times fainter 
than that of canonical short GRBs \citep{2017ApJ...848L..23F}. 
The sub-relativistic top-hat jet model \citep{2017arXiv171005905I}, 
the model of the scattered jet emission by a cocoon \citep{2017arXiv171100243K}, 
the model of the off-axis emission of the structured jet \citep{2017MNRAS.472.4953L, 2017arXiv171203237L,2018ApJ...856L..18M}, 
and the model of the mildly relativistic shock-breakout emission of a cocoon \citep{2017Sci...358.1559K, 2017arXiv171005896G,2018Natur.554..207M}
were proposed to explain the rising X-ray afterglow and weak prompt emission.
According to these models, the viewing angle from the relativistic jet is $\sim$30 deg, 
which is too large in comparison with that of canonical short GRBs.
On the other hand, in the prompt emission of GRB170817A, a soft tail emission following the main pulse was observed \citep{2017ApJ...848L..14G}. 
The soft tail had a blackbody spectrum with $k_{\rm B} T =$ 10.3 $\pm$ 1.5 keV, and the energy flux was $\sim$5$\times 10^{-8}$ erg cm$^{-2}$ s$^{-1}$. 
In the previous observations of canonical short GRBs, 
some GRBs had an extended emission lasting for $\sim$100 s after the short hard pulse \citep{2006ApJ...643..266N}.
GRB050709 was a short hard GRB that was followed 25 s later by a long-soft bump of duration approximately 100 s. The bump was from the same position as that of the short pulse, and the peak flux was 1.53 $\pm$ 0.27 $\times 10^{-8}$ erg cm$^{-2}$ s$^{-1}$ in the 2--10 keV band \citep{2005Natur.437..855V}. 
\citet{2014ApJ...796...13N} reported that a soft X-ray extended emission from BNS mergers was emitted by a mildly relativistic fireball powered by the rotation energy of the Kerr BH via the Blandford-Znajek process.
\citet{2017ApJ...848L..14G} also estimated the 3 $\sigma$ upper limit of extended emission of GRB170817A by GBM as 6.4--6.6$\times 10^{-8}$ erg cm$^{-2}$ s$^{-1}$ with 10-s exposure, which converts to an estimated flux of $\sim$3$\times 10^{-9}$ erg cm$^{-2}$ s$^{-1}$ at a distance of 170 Mpc.
If the soft tail lasts for $\sim$100 s after the pulse emission, 
MAXI/GSC can observe the extended emission of the soft tail.
We consider that the MAXI/GSC observation of the EM counterpart
would contribute to the testing of the X-ray emission model of BNS mergers.

\begin{ack}
This research has made use of the MAXI data provided by RIKEN, JAXA, and the MAXI team. 
This research was supported by JSPS KAKENHI Grant Numbers JP17H06362, 16K05301(HN), 17K05402(MS), and 24684015(KY).
This research made use of data supplied by the UK Swift Science Data Centre at the University of Leicester.
\end{ack}

\appendix
\section{Upper-limit calculation for the source photon counts}
\label{sec:ul}

We estimated the upper limit of the source photon counts for detection with 3 $\sigma$ significance from the background.
In the X-ray observation of an astronomical object, 
the observed net counts $C_{\rm net}$ in the FOV 
contain the source counts $C_{\rm src}$ and the background counts $C_{\rm bg}$: $C_{\rm net} = C_{\rm src} + C_{\rm bg}$.
For N $\sigma$ detection from $C_{\rm bg}$, 
$C_{\rm src}$ should satisfy the following equation: 
\begin{equation}
\label{equ:ul}
C_{\rm src} = C_{\rm net} - C_{\rm bg} > N \sqrt{\sigma_{\rm net}^{2} + \sigma_{\rm bg}^{2}}.
\end{equation}
Based on Poisson statistics for photon counts, the statistical errors are 
$\sigma_{\rm net} = \sqrt{C_{\rm net}}$ and $\sigma_{\rm bg} = \sqrt{C_{\rm bg}}$.
From the solution of the quadratic equation of $C_{\rm net}$,
\begin{equation}
C_{\rm net} > C_{\rm bg} + N \left(\frac{N+\sqrt{8 C_{\rm bg} + N^{2}}}{2}\right),
\end{equation}
$C_{\rm net}$ is described as a function of $C_{\rm bg}$ and $N$.
Then, the upper limit of source photon counts for $N \sigma$ detection is estimated as
\begin{equation}
C_{\rm src}(N,C_{\rm bg}) = N \left(\frac{N+\sqrt{8 C_{\rm bg} + N^{2}}}{2}\right).
\end{equation}

\bibliography{ref}
\bibliographystyle{pasj}

\end{document}